\documentclass[fleqn,twoside]{article}
\usepackage{espcrc2}

\usepackage{graphicx}
\usepackage[figuresright]{rotating}

\newcommand{\AmS}{{\protect\the\textfont2
  A\kern-.1667em\lower.5ex\hbox{M}\kern-.125emS}}

\def\be{\begin{eqnarray}}
\def\ee{\end{eqnarray}}
\def\ttheta{\widetilde{\theta}}

\def\la{\langle}
\def\ra{\rangle}

\def\lb{\lbrack}
\def\rb{\rbrack}

\def\O{{\cal O}}

\title{Improving the locality of the overlap Dirac operator via approximate
solutions of the Ginsparg-Wilson relation}

\author{David H. Adams\address{Instituut-Lorentz for Theoretical Physics, 
        Leiden University, \\ 
        Niels Bohrweg 2, NL-2333 CA Leiden, The Netherlands}
        \thanks{Supported by a Marie Curie fellowship from the European Commission,
contract HPMF-CT-2002-01716}}
       
\begin{document}

\begin{abstract}
We determine a free field hypercubic lattice Dirac operator which is optimally 
close to satisfying the Ginsparg-Wilson relation. Inserting this operator into
the overlap formula, we show that the analytic locality bound on the resulting 
overlap Dirac operator is substantially stronger than in the standard case.
The improvement generally persists in gauge backgrounds when the plaquette variables
are all close to unity.
\vspace{1pc}
\end{abstract}

\maketitle


The overlap Dirac operator provides a formulation of lattice QCD with exactly 
massless quarks \cite{Neu1}. 
Actually there are many overlap operators: for each choice of ultra-local lattice
Dirac operator $D$ an overlap Dirac operator $D_{ov}$ is obtained by inserting $D$
into the overlap formula:
\be
D_{ov}=\frac{m}{a}(1+A(A^*A)^{-1/2})\ ,\ \ A\,\equiv\,D-\frac{m}{a}
\label{1}
\ee
Here $a$ is the lattice spacing and $m$ is a parameter which controls topological
properties of $D_{ov}$ and the number of fermion species described by the 
corresponding lattice fermion action. In the standard overlap operator,  
$D$ is taken to be the Wilson-Dirac operator $D_w\,$.

The locality issue for $D_{ov}$ is nontrivial due to the inverse square root
$(A^*A)^{-1/2}$ in (\ref{1}), which also causes difficulties for numerical 
implementation. To ensure that the lattice theory is in the right universality
class to reproduce continuum physics, $D_{ov}$ should be exponentially-local.
An exponential-locality bound was derived in \cite{L(local)}:
\be
||(A^*A)^{-1/2}(x,y)||\;\le\;K\,e^{-\ttheta\,|x-y|/a}
\label{2}
\ee
where $|x-y|\equiv\sum_{\mu}|x_{\mu}-y_{\mu}|$. Set $l$ to be the maximum distance 
in units of lattice spacing between lattice sites coupled by $D$ as measured by 
this norm (e.g. $l=1$ for the Wilson-Dirac operator).  Then the decay constant in
(\ref{2}) is
\be
\ttheta=\theta/2l
\label{8}
\ee
with $\theta$ determined by the {\em condition number} $C$:
\be
\theta=\log\Big(\frac{\sqrt{C}+1}{\sqrt{C}-1}\Big)\ ,\ \ 
C=\frac{\lambda_{max}(A^*A)}{\lambda_{min}(A^*A)}
\label{4}
\ee
$\lambda_{max}$ ($\lambda_{min}$) are the maximum (minimum) eigenvalues.
Moreover $K=\lambda_{min}(A^*A)^{-1/2}$
(this and (\ref{8})--(\ref{4}) were not written out explicitly in 
\cite{L(local)} but follow easily from the formulae derived there).
To obtain a locality bound independent of the gauge field, gauge field-independent 
bounds $0<u\le A^*A \le v<\infty$
are required; then $\lambda_{min}$ and $\lambda_{max}$ can be replaced by $u$ and $v$,
respectively, in the preceding. An upper bound $v$ can easily be derived via
triangle inequalities. However, to get a nonzero lower bound $u$ some restriction on
the lattice gauge fields is generally required, since gauge backgrounds in which
$A$ has zero-modes do exist in general (reflecting topological properties of $D_{ov}$).
In the standard Wilson-Dirac case this can be done by imposing an {\em admissibility 
condition} \cite{L(local),Neu(bound),DA(bound)}
\be
||1-U(p)||<\epsilon\quad\forall\,\mbox{plaquette $p$}.
\label{7}
\ee
The currently sharpest lower bound is $1-6(2+\sqrt{2})\epsilon\le A_w^*A_w$ 
(in 4 dimensions and with $m=1$) \cite{Neu(bound)}.

In \cite{Biet(EJP)} the possibility of improving locality and convergence properties 
of the overlap Dirac operator by taking the input $D$ to be an approximate solution 
of the Ginsparg-Wilson (GW) relation \cite{GW} was pointed out. 
This is based on the observation that if $D$ is an exact GW solution, i.e. 
$\gamma_5D+D\gamma_5=\frac{a}{m}D\gamma_5D$, then the overlap formula simply gives 
this operator back again: $D_{ov}=D$. (To see this, note that the GW relation
is equivalent to $A^*A=(\frac{m}{a})^2$ where $A\equiv D-\frac{m}{a}$.)
It is known that ultra-local operators cannot exactly satisfy the GW relation
\cite{nogo}; but approximate solutions are possible, and for these
the preceding observation implies $D_{ov}\approx D$, indicating that $D_{ov}$ 
is ``close to being ultra-local''. This heuristic reasoning needs to be treated 
with caution though, because if $D$ is very close to satisfying the GW relation
then it must be ``almost non-ultralocal''. In this note we attempt to get a more
precise analytic understanding of the situation by considering the improvement that
can be achieved in the locality bound (\ref{2}). If $D$ is an approximate
GW solution, i.e. $A^*A\approx(\frac{m}{a})^2$, then 
$\lambda_{min}\approx\lambda_{max}\approx(\frac{m}{a})^2$, hence the condition number 
$C$ is close to unity and consequently the parameter $\theta$ in (\ref{4}) is very
large. However, the effect of the the increase in $\theta$ in the decay constant
(\ref{8}) must be at least partly offset by an increase in $l$,
since the closer $D$ is to satisfying the GW relation the longer its range must be.
Therefore, to improve the locality bound on the overlap Dirac operator, we are led to 
look for short range approximate solutions to the GW relation to use as input in the
overlap formula.


A natural arena in which to look for such operators
is the class of hypercubic operators. These are the lattice Dirac operators which
couple sites within the same lattice hypercube. For such operators the parameter
$l$ in (\ref{8}) equals the spacetime dimension $d$, so in 4 dimensions
an increase in $\theta$ by more than a factor of 4 is required in order to achieve
a locality bound that is stronger than the one for the standard overlap operator. 
We begin by considering the free field case. Define the hermitian operators
$S_{\mu}=\frac{1}{2i}(T_{+\mu}-T_{\-\mu})$ and $C_{\nu}=\frac{1}{2}(T_{+\nu}+T_{-\nu})$
where $T_{+\mu}$ ($T_{-\mu}$) are the usual forward (backward) parallel transport 
operators. Then a general free field hypercubic operator can be written in terms of 
the free field $S_{\mu}$'s and $C_{\nu}$'s as \cite{AB}
\be
D=\frac{1}{a}(\gamma^{\mu}\rho_{\mu}+\lambda)
\label{9}
\ee
$-i\rho_{\mu}\!=\!S_{\mu}\sum_{p=1}^d\!2^p\kappa_p\!
\sum_{\nu_2<\cdots<\nu_p\,,\,\nu_j\ne\mu\forall j}C_{\nu_2}\cdots C_{\nu_p}$
and $\lambda=\sum_{p=0}^d2^p\lambda_p\sum_{\nu_1<\cdots<\nu_p}
C_{\nu_1}\cdots C_{\nu_p}$
Here $\kappa_1,\dots,\kappa_d$ and $\lambda_0,\dots,\lambda_d$ are coupling parameters; 
they are the most general couplings allowed by the lattice symmetries. The requirements
of correct formal continuum limit ($D\to\gamma^{\mu}\partial_{\mu}$) and vanishing bare
mass are equivalent to the constraints
$\sum_{p=1}^{d}2^p{\textstyle \left\lb{d-1 \atop p-1}\right\rb}
\kappa_p=1$ and $\sum_{p=0}^{d}2^p
{\textstyle \left\lb{d \atop p}\right\rb}\lambda_p=0$.
These can be used to eliminate $\kappa_1$ and $\lambda_0$; then $D$ contains $2d-1$
free parameters $\kappa_2,\dots,\kappa_d;\lambda_1,\dots,\lambda_d$. (The Wilson-Dirac 
operator, which contains just one free parameter --the Wilson parameter $r$ -- is 
obtained by setting $\kappa_1=1/2$, $\lambda_0=dr$, $\lambda_1=-r/2$ and 
$\kappa_p=\lambda_p=0$ for $p\ge2$.)

A number of free field hypercubic operators which approximately satisfy the GW 
relation are already known and have been discussed in \cite{Biet(EJP)}:
the truncation of the standard overlap operator, the truncated Fixed Point (FP) 
operator (the FP operator is a GW solution which is known explicitly only in the free 
field case \cite{FP}), and the ``GW-improved'' operator obtained by inserting
the truncated FP operator into the overlap formula and then truncating again. 
However, one can attempt
to do better still by considering the condition number (\ref{4}) as a function 
$C=C(\kappa_2,\dots,\kappa_d;\lambda_1,\dots,\lambda_d)$ on the
parameter space of free field hypercubic operators
and finding the point(s) at which $C$ has a minimum. We did this numerically, 
using the expression $C={\textstyle {max \atop k}}\;\{A^*A(k)\}
\Big/\,{\textstyle {min \atop k}}\;\{A^*A(k)\}$
where $A^*A(k)$ is the free field momentum representation of $A^*A$.
For concreteness we fix the spacetime 
dimension to be $d=4$ and set $m=1$ in (\ref{1}), 
i.e. $aA\equiv aD-1$. The lattice spacing $a$ drops out in the expression for $C$.
When calculating a minimum of $C(\kappa_2,\dots,\lambda_4)$ 
numerically one must specify an initial point (actually two initial points for our 
Mathematica calculation). We tried various choices of starting
points corresponding to the truncated FP operator and other free field hypercubic
operators considered in \cite{Biet(EJP)}. In each case the the numerical calculation
converged to the same minimum of $C(\kappa_2,\dots,\lambda_4)$ at the point in 
parameter space listed in the last column of Table~\ref{T1}.
\begin{table}
\begin{flushright}
\begin{tabular}{|c|c|c|}
\hline
& Trunc. FP & Optimal \\
\hline
$\kappa_2$     & 0.03208  & 0.03230  \\
$\kappa_3$     & 0.01106  & 0.01107  \\
$\kappa_4$     & 0.00475  & 0.00613  \\
$\lambda_1$    & -0.06076  & -0.06191 \\
$\lambda_2$    & -0.03004  & -0.03027 \\
$\lambda_3$    & -0.01597  & -0.01576 \\
$\lambda_4$    & -0.00843  & -0.00778 \\
\hline
\end{tabular}
\end{flushright}
\vspace*{-6mm}
\caption{\it{The coupling parameters of the new ``optimal'' free field hypercubic 
operator. For comparison the parameters of the truncated FP operator are also 
listed.}}
\label{T1}
\vspace*{-7mm}
\end{table}
Thus we have found a new free field
hypercubic operator which is optimally close to satisfying the GW relation, at least
compared to other operators in a neighbourhood of the truncated
FP operator. 

In Table~\ref{T2} we list $\lambda_{min}\,$, $\lambda_{max}$ and the condition number
$C$ for the new operator, as well as the decay constant $\ttheta$ in the locality
bound (\ref{2}) for the corresponding overlap Dirac operator, and compare with
the values for other operators. The $\ttheta$ for the new 
operator is significantly larger than for the previously considered free field
hypercubic operators, and is larger than the Wilson-Dirac $\ttheta$ by more than a 
factor of 3.
\begin{table}
\begin{flushright}
\begin{tabular}{|c|c|c|c|c|}
\hline
& $\lambda_{min}$ & $\lambda_{max}$  & $C$ & $\ttheta$ \\
\hline
Wilson-Dirac & 1 & 49 & 49    & 0.14 \\
Trunc. FP    & 0.919 & 1.023 & 1.113 & 0.45 \\ 
GW-improved  & 0.932 & 1.032 & 1.107 & 0.46 \\
Optimal      & 0.938 & 1.005 & 1.072 & 0.51 \\
\hline
\end{tabular}
\end{flushright}
\vspace*{-1mm}
\caption{\it{$\lambda_{min}(A^*A)$, $\lambda_{max}(A^*A)$, $C$ and $\ttheta$ for the 
free field Wilson-Dirac operator, the truncated FP and its GW-improved version,
and the new ``optimal'' hypercubic operator.}}
\label{T2}
\vspace*{-9mm}
\end{table}

Gauged hypercubic operators can be built up in various ways from free field operators.
We now point out that if the free field operator is 
a good approximate GW solution then the same is generally true for the gauged operator 
when the gauge field satisfies the admissibility condition (\ref{7}) with small
$\epsilon$. General ultra-local gauged lattice Dirac operators are polynomials in the 
parallel transporters $T_{\pm\mu}$. Since these can be expressed in terms of the
$S_{\mu}\,,\,C_{\nu}$ defined earlier, $A^*A=A^*A(S_{\mu},C_{\nu})$. 
Inserting the spectral decompositions
$S_{\mu}=\sum_{\alpha_{\mu}}s_{\mu\alpha_{\mu}}P_{\mu\alpha_{\mu}}\,$, 
$C_{\nu}=\sum_{\beta_{\nu}}c_{\nu\beta_{\nu}}P_{\nu\beta_{\nu}}$ 
(where $P_{\mu\alpha_{\mu}}\,$, $Q_{\nu\beta_{\nu}}$ are the projections onto the 
eigenspaces with eigenvalues $s_{\mu\alpha_{\mu}}\,$, $c_{\nu\beta_{\nu}}\,$, 
respectively), and using the fact that in any gauge background
$s_{\mu\alpha_{\mu}},c_{\nu\beta_{\nu}}\in[-1,1]$,
one can make a connection with $A^*A_{free}$ and derive bounds 
\be
\lambda_{min,f}-\lambda_{min,f}(C_f-1)K_--O(\epsilon)
\,\le\,A^*A && \nonumber \\
\le\,\lambda_{max,f}+\lambda_{min,f}(C_f-1)K_-
+O(\epsilon) && \label{30}
\ee
(the derivation will be given elsewhere)
where the subscript ``$f$'' refers to the free field quantities.
Here $K_-={max \atop ||\psi||=1}\,|\sum_{\sigma(-)}\la\psi,\O_{\sigma}\psi\ra|$
where $\O_{\alpha_1\cdots\beta_4}\equiv P_{1\alpha_1}\cdots Q_{4\beta_4}$ and the sum 
is restricted to the $\sigma\!=\!\{\alpha_1,\dots,\beta_4\}$ for which 
$\la\psi,\O_{\sigma}\psi\ra$ is negative.
Since $\sum_{\sigma(+)}\la\psi,\O_{\sigma}\psi\ra+
\sum_{\sigma(-)}\la\psi,\O_{\sigma}\psi\ra=1$ for all unit norm $\psi$, and 
$|\la\psi,\O_{\sigma}\psi\ra|\le1$ $\forall\sigma$, we can expect $K_-$ to be of
order 1 in typical gauge backgrounds. Then, if the free field operator is a good
approximate GW solution (i.e. $C_f\approx1$), the terms with $K_-$ are very small
in (\ref{30}). The bounds then imply that
the condition number of the gauged operator is close to the free field condition
number when $\epsilon$ is small. Finally we mention that hypercubic and other short 
range lattice Dirac operators which are reasonable 
approximate GW solutions in equilibrium gauge backgrounds have already been found
in numerical work; see \cite{Biet(NPB)} and the ref.'s therein.

I thank Wolfgang Bietenholz for discussions.

\end{document}